\documentclass[sigconf]{acmart}
\setcopyright{none}
\renewcommand\footnotetextcopyrightpermission[1]{}
\usepackage{makecell}
\usepackage{multirow}
\usepackage[table,xcdraw]{xcolor}

\usepackage{booktabs}
\AtBeginDocument{%
  }

\begin{document}

\title{Can Asymmetric Tile Buffering Be Beneficial?}



\author{%
  Chengyue Wang$^{1}$,
  Wesley Pang$^{2}$,
  Xinrui Wu$^{1}$,
  Gregory Jun$^{2}$,
  Luis Romero$^{3}$,
  Endri Taka$^{4}$,
  Diana Marculescu$^{4}$,
  Tony Nowatzki$^{1}$,
  Pranathi Vasireddy$^{5}$,
  Joseph Melber$^{5}$,
  Deming Chen$^{2}$,
  Jason Cong$^{1}$%
}
\affiliation{%
  \institution{$^{1}$UCLA \quad
               $^{2}$UIUC \quad
               $^{3}$EPFL \quad
               $^{4}$UT Austin \quad
               $^{5}$AMD}%
    \country{}
}
\email{chengyue@ucla.edu, qpang2@illinois.edu, cong@cs.ucla.edu}



\begin{abstract}


General matrix multiplication (GEMM) is the computational backbone of modern AI workloads, and its efficiency is critically dependent on effective tiling strategies. Conventional approaches employ \emph{symmetric tile buffering}, where the buffered tile size of the input $A$ along the dimension $M$ matches the output tile size of $C$.

In this paper, we introduce \emph{asymmetric tile buffering} (ATB), a simple but powerful technique that decouples the buffered tile dimensions of the input and output operands. 
We show, for the first time, that ATB is both practical and highly beneficial. 
To explain this effect, we develop a performance model that incorporates both the benefits of ATB (higher arithmetic intensity) and its overheads (higher kernel switching costs), providing insight into how to select effective ATB tiling factors. 
As a case study, we apply ATB to AMD’s latest XDNA2™ AI Engine (AIE), achieving up to a 4.54$\times$ speedup, from 4.8 to 24.6 TFLOPS on mixed-precision BFP16–BF16 GEMM, establishing a new performance record for XDNA2™ AIE.

\end{abstract}




\maketitle

\section{Introduction}

The rapid progress of artificial intelligence (AI) has placed unprecedented demands on efficient GEMM computations. In GEMM, we perform matrix multiplication $C = A \times B$, where $A \in \mathbb{R}^{M \times K}$, $B \in \mathbb{R}^{K \times N}$, and $C \in \mathbb{R}^{M \times N}$, with $M$, $N$, and $K$ denoting the number of rows of $A$, the number of columns of $B$, and the reduction dimension, respectively. GEMM efficiency relies critically on effective tiling strategies, which increase arithmetic intensity by reusing data to reduce total data movement to and from memory. As a result, tiling has become a standard practice across CPUs, GPUs, and domain-specific accelerators, enabling high performance while respecting tight memory-bandwidth constraints.

\begin{figure}[t]
  \centering
\includegraphics[width=\linewidth]{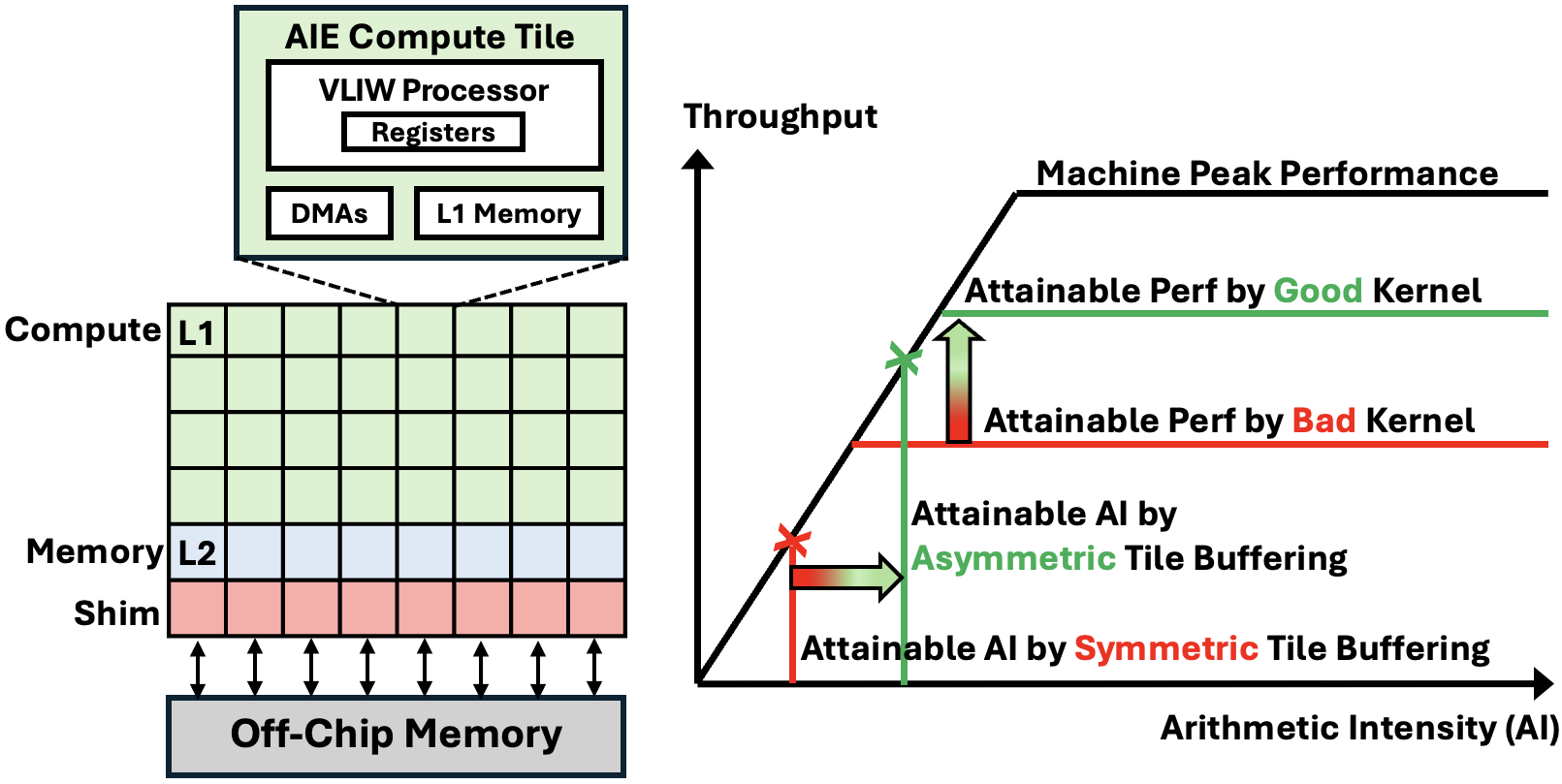}
\caption{Architecture and Roofline: AMD AIE XDNA2™ spatial architecture (left) and the roofline view illustrating the performance impact of kernel and tiling strategies (right).}
\label{fig:aie_roofline}
\end{figure}

Conventional practice is to employ \emph{symmetric tile buffering}, in which the buffered tile size of input $A$ along the $M$ dimension and input $B$ along the $N$ dimension are the same as the output tile size of $C$. This uniform mapping, defined by the three parameters $(T_M, T_K, T_N)$, simplifies scheduling and implementation and is widely adopted in CPU, GPU, and NPU software libraries, as well as hardware accelerator designs~\cite{BLIS1, Thakkar_CUTLASS_2023, zhuang2023charmcomposingheterogeneousaccelerators, rsn}. Despite its ubiquity, symmetric buffering increases buffer requirements and leaves arithmetic intensity below the theoretical maximum (Fig.~\ref{fig:aie_roofline}, right), a limitation that is particularly restrictive when targeting narrow memories such as L1 caches or scratchpads.

However, we show that this symmetry is unnecessary for tiling. In principle, the buffer lifetime of input $A$ is only as long as it takes to accumulate a single row of $C$; once that row is complete, the corresponding row of $A$ can be safely evicted. This observation implies that the input and output tile buffers do not need to share the same $M$ dimension, opening the opportunity to reduce the buffering requirements for $A$ and allocate more capacity to $B$ and $C$.
This insight raises the central question of this work: \emph{When and how can asymmetric tile buffering be beneficial?}
To answer it:
\begin{itemize}
\item We introduce ATB, a general tiling strategy that decouples the buffered $M$ dimensions of $A$ and $C$, reducing input buffer pressure and increasing arithmetic intensity.
\item We develop an analytical performance model that captures both the benefits of ATB (higher reuse and arithmetic intensity) and its costs (more kernel switching overheads), providing principled guidance for selecting tiling factors.
\item We present the first case study applying ATB to AMD NPU, achieving a speedup of up to 4.54$\times$ over the state of the art.
\item We analyze ATB’s performance on AIE and show that it improves GEMM throughput by up to 40\% even over highly optimized symmetric kernels.
\end{itemize}

\section{Background}

\subsection{Reconfigurable XDNA2™ Architecture}

Fig.~\ref{fig:aie_roofline} (left) shows the latest AMD XDNA2™ NPU architecture, integrated in Strix ~\cite{amd-ryzen-ai-9-hx-370} and Kraken ~\cite{amd-ryzen-ai-5-340} Point laptops. It features an AIE array consisting of 32 compute cores, eight shim cores, and eight memory cores. Those cores are reconfigurable blocks connected with FPGA-like adaptable interconnects~\cite{amd_adaptive_overview}.
Each compute core integrates a VLIW processor, 64 KB of local L1 memory, DMA engines, register files, and two input and two output stream channels. Inter-core connections are reconfigurable through stream channels, enabling flexible communication patterns such as broadcasting from a single producer core to multiple consumer cores. Each memory core provides 512 KB of L2 memory along with six input and six output stream channels. 
To access the flexible compute and memory hierarchy, toolkits such as IRON~\cite{fccmIRON} provide explicit control over data movement, allowing developers to configure stream interfaces and data movement across cores in Python while linking C++ AIE kernels for programming the VLIW processor.

\subsection{Challenges of Data Movement and Tiling}
\label{sec:cha_of_data_movement}
Arithmetic intensity measures the ratio of computation to data movement and is critical for overcoming memory bottlenecks (Fig.~\ref{fig:aie_roofline}, right) \cite{roofline, mindgap}. 
For GEMM, arithmetic intensity increases with tile size. 
Each AIE compute core provides only two input streams, typically used for operands $A$ and $B$, leaving no channel to bring partial sums back for further accumulation. Moreover, memory cores lack accumulator units, so partial sums cannot be stored in L2 or reloaded to the compute cores.
As a result, each output tile of $C$ must be fully accumulated in L1 before being written out, and the L2 output tile size exactly equals the aggregate of L1 tiles across the used compute cores. 
For example, if each core computes a $64\times 64$ BF16 tile (8~KB), then 32 cores collectively produce 256~KB of output in L1, so the L2 output tile size is also 256~KB. 
The L1 output tile size is therefore critical: it directly determines the L2 output tile size, which directly governs the reuse of off-chip input data.

Given this, asymmetric tile buffering in L1 memory offers a promising way to increase arithmetic intensity and improve off-chip data reuse.
However, prior AIE studies 
\cite{zhuang2023autommenergyefficientmultidatatypematrix, AMA_FPL_2024, Versal_vs_Stratix_FCCM_2024, zhuang2023charmcomposingheterogeneousaccelerators, taka2023maxevamaximizingefficiencymatrix, mhatre2025gamahighperformancegemmacceleration} 
have adopted only symmetric tile buffering at both L1 and L2, with the tiling space defined solely by the three parameters ($T_M$, $T_K$, $T_N$) at each memory level. 
Although simulation-based design space exploration studies \cite{mei2020zigzagmemorycentricrapiddnn, gilbert2024looptree} have explored the idea of asymmetric tile buffering, they neither analyze its overheads on real hardware nor its impact on kernel performance. 
Moreover, to the best of our knowledge, CPU and GPU implementations also rarely adopt such strategies, due to both hardware constraints and the complexity of managing data movement across multiple memory levels. For example, widely used GEMM frameworks such as BLIS \cite{BLIS1} and CUTLASS \cite{Thakkar_CUTLASS_2023} also rely on symmetric tile buffering in L1 caches.

\subsection{Challenges of Kernel Design}

As shown in Fig.~\ref{fig:aie_roofline} (right), an efficient kernel design drives the bound of attainable performance.
Inside a kernel, the fundamental unit of computation is the \emph{microkernel}: a fully unrolled inner loop that performs fused multiply–accumulate (MAC) operations on small tiles of data held in registers. 
Running on a 7-way VLIW processor, AIE microkernel performance depends on how efficiently the \texttt{VMAC} instruction is scheduled at compile time.

While most prior AIE studies \cite{zhuang2023autommenergyefficientmultidatatypematrix, AMA_FPL_2024, Versal_vs_Stratix_FCCM_2024, zhuang2023charmcomposingheterogeneousaccelerators, taka2023maxevamaximizingefficiencymatrix, mhatre2025gamahighperformancegemmacceleration} 
focus primarily on data movement between L1, L2, and off-chip memory, they generally lack a systematic discussion of AIE microkernel optimization.
Additionally, although microkernels can be optimized with compiler pragmas such as \texttt{chess::unroll\_loop} in the Chess compiler for AIE, analytical modeling of the underlying execution is essential to guide programmers in their effective use of pragmas. 
Furthermore, mixed precision complicates instruction pipelines and makes it difficult for the compiler to optimize without manual code refactoring. This paper presents an analytical model to guide manual kernel optimization.

Between kernels, additional performance challenges arise from execution gaps introduced during kernel switching. These gaps result from inherent overheads such as kernel initialization, pipeline resets, and synchronization delays. They accumulate across the many kernels invoked in a tiled GEMM, widening the difference between microkernel-level attainable performance and core-level attainable performance. 


\subsection{BFP16 Data Type}
XDNA2™ introduces native support for the Block Floating Point (BFP16) format~\cite{amd-bfp16-quark}, a quantized precision that balances efficiency and accuracy. Unlike BF16, where each value carries its own exponent, BFP16 shares one 8-bit exponent across a block of 8 values, each storing only an 8-bit mantissa. This packs 8 elements into 9 bytes, represented in AIE as \texttt{bfp16ebs8}~\cite{amd-aie-api-mmul}.
Each compute core executes BFP16 on a dedicated $8\times8\times8$ MAC unit, sustaining 512 MACs per cycle. 
A \texttt{VMAC} instruction multiplies two $8\times8$ BFP16 vector registers and accumulates into an $8\times8$ FP32 accumulator register. At 1.8 GHz \cite{AMD-xdna-driver}, this delivers 1.84 TFLOPS per core and 58.8 TFLOPS across the whole 32-core array. However, the current best mixed-precision BF16-BFP16 GEMM result on an AMD Strix Point laptop is only 4.8~TFLOPS~\cite{mlir-aie-matmul}.



\section{Modeling Arithmetic Intensity}

\begin{figure}[t]
\vspace{-10pt}
  \centering
  \includegraphics[width=\linewidth]{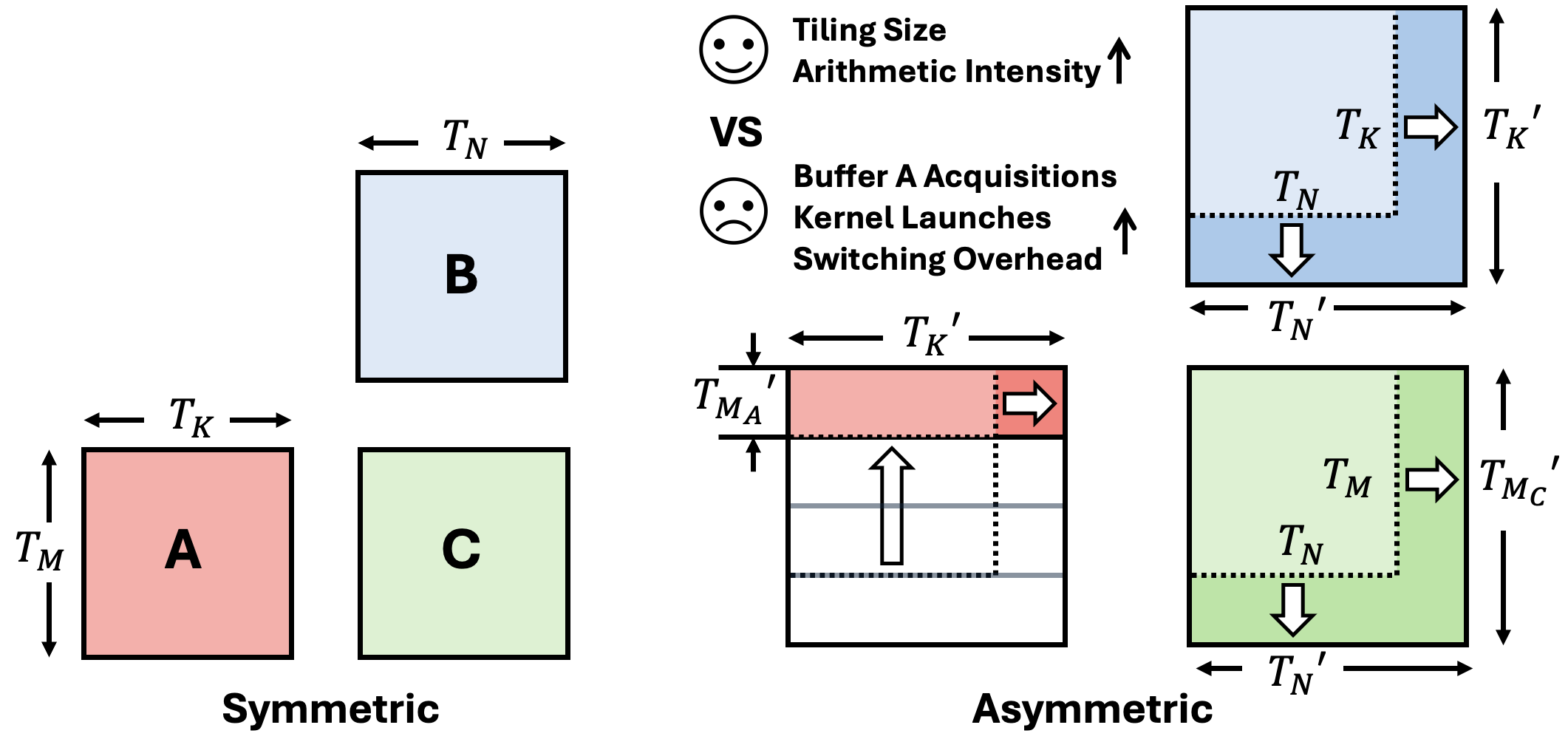}
  \vspace{-10pt}
\caption{Comparison of symmetric (left) and asymmetric (right) tile buffering. Asymmetric tiles ($T_{M_A}$, $T_{M_C}$, $T_K$, $T_N$) increase arithmetic intensity but also switching overhead.}
~\label{fig:assym_tiling}
    \vspace{-10pt}

\end{figure}

In GEMM, symmetric tile buffering uses $(T_M, T_K, T_N)$ to define tile sizes for $A$, $B$, and $C$, where 
$A_{\text{tile}} \in \mathbb{R}^{T_M \times T_K}$, 
$B_{\text{tile}} \in \mathbb{R}^{T_K \times T_N}$, 
and $C_{\text{tile}} \in \mathbb{R}^{T_M \times T_N}$.
Within each tile, consider the canonical loop order 
\texttt{for m in $T_M$ \{ for n in $T_N$ \{ for k in $T_K$ \{\}\}\}}. 
For a fixed $m$, the row $A_{\text{tile}}[m,:]$ is reused across all $n$ until the corresponding row $C_{\text{tile}}[m,:]$ has been fully accumulated, after which that row of $A$ is no longer needed. In contrast, each row of $C_{\text{tile}}$ must remain live for the entire reduction over $K$ (full problem or next-level tile size).
Because the lifetimes of $A$ and $C$ differ along the dimension $M$, their buffered extents do not need to be equal. This enables \emph{asymmetric tile buffering}, parameterized by $(T_{M_A}, T_{M_C}, T_K, T_N)$, where $T_{M_A}$ buffers rows of $A$ and $T_{M_C}$ buffers rows of $C$ (Fig.~\ref{fig:assym_tiling}). In this scheme,
$A_{\text{tile}} \in \mathbb{R}^{T_{M_A} \times T_K}$, 
$B_{\text{tile}} \in \mathbb{R}^{T_K \times T_N}$, and 
$C_{\text{tile}} \in \mathbb{R}^{T_{M_C} \times T_N}$ ($T_{M_C}\geq T_{M_A}$).

\subsection{Arithmetic Intensity with ATB }

Let $a,b,c>0$ be per-element byte costs (e.g., 2 for BF16 and 1.25 for BFP16) for $A$, $B$, and $C$, and define the asymmetry ratio
$\rho = T_{M_C}/T_{M_A}$ ($\rho \geq 1$). We consider an output-stationary schedule: each
output tile is fully iterated $N_{\text{iterk}} = K/T_K$ times (reading $A,B$ each time) and
written once at the end, where $K$ is the problem size along the reduction
dimension of the GEMM.
For a tile of size $(T_{M_A},T_{M_C},T_K,T_N)$, the arithmetic intensity is  

\begin{align*} \\[-25pt]
\mathrm{AI}_\rho &= \frac{\text{Operations}}{\text{Accesses}}
= \frac{\text{Operations per output tile}}{\text{Accesses per output tile}} \\[4pt]
&= \frac{2\,T_{M_C}T_KT_N\,N_{\text{iterk}}}{a\,T_{M_C}T_K N_{\text{iterk}} + b\,T_KT_N N_{\text{iterk}} + c\,T_{M_C}T_N} \\[4pt]
&= \frac{2\,K\,T_{M_C}T_N}{a\,T_{M_C}K + b\,KT_N + c\,T_{M_C}T_N} = \frac{2}{\tfrac{a}{T_N} + \tfrac{b}{T_{M_C}} + \tfrac{c}{K}},
\end{align*}
which is bounded by the available buffer capacity $S$, assuming double buffering of $A$, $B$ and single buffering of $C$:
\begin{equation}
\label{eq:constraint}
\tfrac{2a}{\rho}\,T_{M_C}T_K \;+\; 2b\,T_KT_N \;+\; c\,T_{M_C}T_N \;\le\; S.
\end{equation}
As shown above, $\mathrm{AI}_\rho$ increases monotonically with $T_{M_C}$ and $T_N$. Importantly, $\mathrm{AI}_\rho$ does not depend explicitly on $T_K$ or $\rho$; their effects appear only through the buffer constraint (Eq.~\ref{eq:constraint}). 
The smaller $T_K$ and the larger $\rho$ reduce the buffer cost associated with $T_{M_C}$ and $T_N$. This frees more buffer space, allowing larger $(T_{M_C}, T_N)$ values to be used, which in turn increases $\mathrm{AI}_\rho$.
The optimal strategy to maximize $\mathrm{AI}_\rho$ is to choose \textbf{$T_K$ as small as possible} and \textbf{$\rho$ as large as possible}. In the idealized case, this corresponds to $T_K = 1$ and $T_{M_A}=1$, with $\rho = T_{M_C}$.

\subsection{From Single Core to Whole Array}

In XDNA2™, computation is organized in three levels of tiling: from off-chip memory to L2, from L2 to L1, and from L1 to registers. As discussed in Section~\ref{sec:cha_of_data_movement}, the L2 output tile is the aggregation of the L1 output tiles produced by all compute cores. From L2 to L1, one input tile in a memory core can be broadcast to multiple compute cores; specifically, L2 $A$ tiles are broadcast across columns, and L2 $B$ tiles are broadcast across rows. We do not apply asymmetric tiling at the L2 level, since the L2 tile size is limited by the L1 tile size and the available L2 memory capacity is sufficient to hold symmetric tiles. For an array of $N_{\text{aie\_row}}=4$ rows and $N_{\text{aie\_col}}=8$ columns, the effective L2 tile sizes are
\[
T^{L2}_M = N_{\text{aie\_row}} \cdot T^{L1}_{M_C} = 4 \, T^{L1}_{M_C}, \hspace{6pt} 
T^{L2}_N = N_{\text{aie\_col}} \cdot T^{L1}_N = 8 \, T^{L1}_N, \hspace{6pt}
T^{L2}_K = T^{L1}_K.
\]
The modeling of arithmetic intensity for data moved from off-chip for the whole array is therefore:
\begin{equation}
\mathrm{AI}_{\text{array}}
= \frac{2}{\tfrac{a}{T^{L2}_N} + \tfrac{b}{T^{L2}_M} + \tfrac{c}{K}} 
= \frac{2}{\tfrac{a}{8 \, T^{L1}_N} + \tfrac{b}{4 \, T^{L1}_{M_C}} + \tfrac{c}{K}},
\end{equation}
which is constrained by the L1 capacity $S_{L1}=63\,\mathrm{KB}$ per core:
\begin{equation}
\label{eq:constraint_whole}
\tfrac{2a}{\rho}\,T^{L1}_{M_C}T^{L1}_K \;+\; 2b\,T^{L1}_KT^{L1}_N \;+\; c\,T^{L1}_{M_C}T^{L1}_N \;\le\; S_{L1} = 63\text{ KB}.
\end{equation}
Similar to the analysis of $\mathrm{AI}_\rho$, the optimal strategy to maximize 
$\mathrm{AI}_{\text{array}}$ is to minimize $T_K^{L1}$ and maximize $\rho$.

\subsection{Limits of AI-Only Modeling}

The arithmetic intensity model predicts that performance is maximized by minimizing $T_K$ and maximizing $\rho$. 
In practice, this overlooks microkernel execution: a very small $T_K$ introduces bubbles in the compute pipeline and inflates prolog/epilog costs. 
To capture these effects, we must incorporate \emph{compute core efficiency}, which itself depends on the tiling factors. 
We refine this in the next section~\ref{sec:microkernel_model}, specifically focusing on AIE's VLIW cores as a case study.

\section{Modeling Instruction-Level Parallelism}

A key characteristic of the VLIW architecture is its deterministic instruction schedule, which is fixed statically at compile time. This property enables performance to be estimated directly from the instruction sequence itself. While runtime hazards such as memory stalls or lock acquisition can introduce variability, the static instruction schedule provides a clear representation of the kernel's peak potential for our analysis.
The performance of a compute core is evaluated by how closely its achieved throughput approaches the theoretical maximum of the hardware, which can be quantified by:
\begin{equation} \text{Efficiency} = \frac{\text{Achieved Throughput (MACs/cycle)}}{\text{Peak Throughput (MACs/cycle)}}. \end{equation} 
The key challenge in kernel design is to exploit instruction-level parallelism (ILP) to hide \emph{data movement latency between L1 memory and registers} and maximize the \emph{issue rate of \texttt{VMAC} instructions}. 
For a GEMM kernel, 
the most critical VLIW slots are: 
\begin{itemize} 
\item \textbf{1 Vector Arithmetic Unit:} Executes the \texttt{VMAC} instruction. 
\item \textbf{2 Vector Load Units:} Load vector operand from L1 memory. 
\item \textbf{1 Vector Store Unit:} Writes the result block back to memory. 
\end{itemize} 
From an instruction-scheduling perspective, maximizing efficiency corresponds to maximizing utilization of the vector arithmetic unit. Thus, we evaluate kernel schedules using the ratio of issued \texttt{VMAC} instructions to the total number of cycles in the code region of interest, typically the microkernel.




\subsection{Accumulation Chain in Microkernel }

In general, instruction scheduling of arbitrary DAGs under resource constraints is NP-hard \cite{Gillies-Liu-ANDOR-1990} and admits no closed-form solution. 
This section narrows the scope to the GEMM microkernel. 
Within this setting, the instruction schedule can be well approximated using simple analytical bounds to guide microkernel optimization.

The basic unit in our model is the \emph{accumulation chain}, which represents the sequence of instructions required to update one accumulator operand of a \texttt{VMAC} instruction without spilling the accumulator register. 
All instructions are pipelined with a throughput of one per cycle, but may have varying latency. 
We decompose the instruction scheduling into three phases:

\begin{figure}[t]
  \centering
  \vspace{-10pt}
  \includegraphics[width=0.8\linewidth]{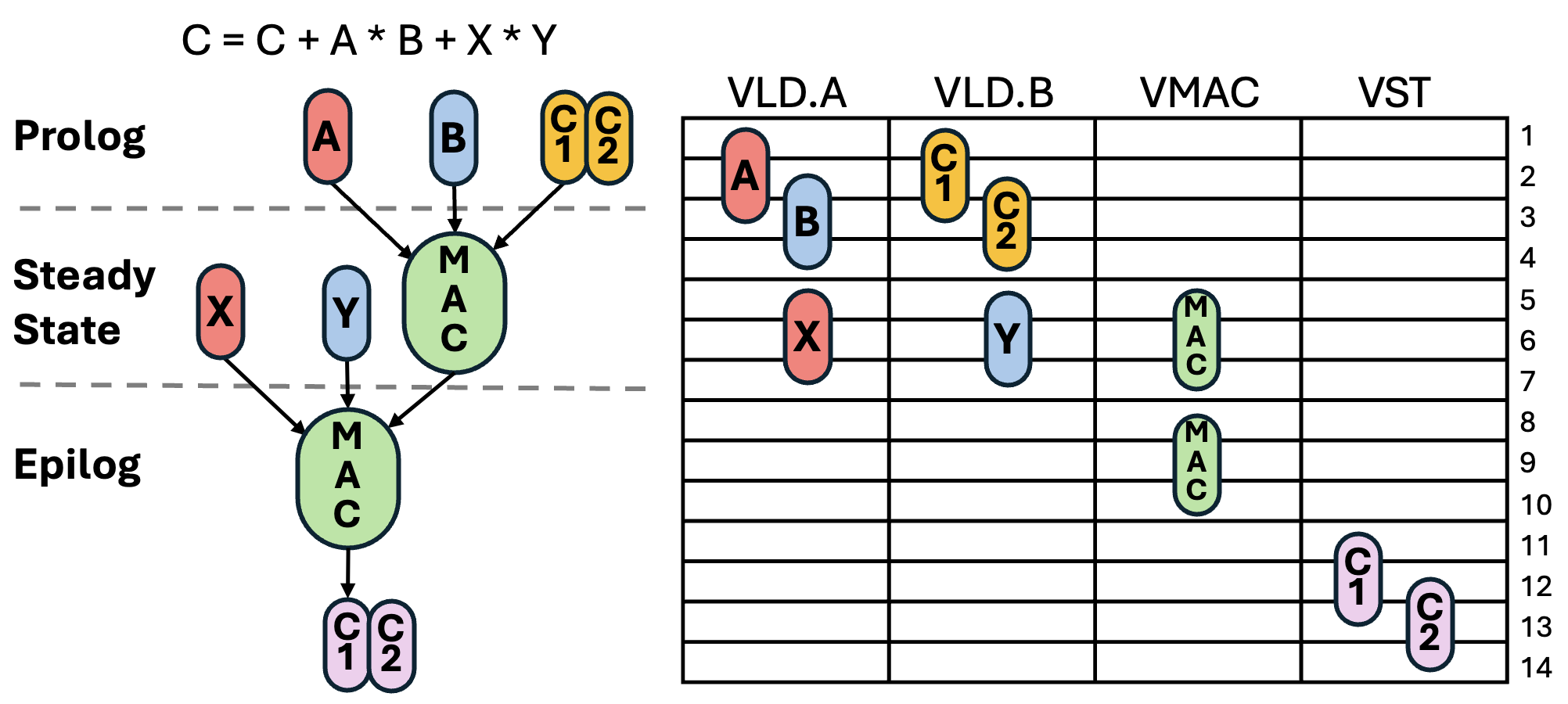}
  \vspace{-10pt}
    \caption{Example schedule of an accumulation chain. 
    Assume 3-cycle latency per instruction, and loading one input takes 1 instruction, accumulator load/store takes 2.}
    \label{fig:accum_chain}
    \vspace{-10pt}
\end{figure}

\setlength{\itemindent}{0pt}
\setlength{\leftmargini}{1em}
\begin{itemize}
\item \textbf{Prolog:} The period before the first \texttt{VMAC} instruction can be issued, during which operand registers are loaded (e.g., loads of $A$, $B$, and $C$ in Fig.~\ref{fig:accum_chain}).
Assume there are $M$ instruction types, each with latency $\ell_i$ and a count $n_i$, to be issued on $U_{\text{LD}}$ vector load slots. 
The minimum time for all operands to become ready is
\vspace{-3pt}
\begin{equation}
T_{\text{prolog}} \;\geq\; T_{\text{load}} = \max_{i=1,\dots,M} \left( \ell_{(i)} + \Big\lceil \tfrac{S_{(i)}}{U_{\text{LD}}} \Big\rceil - 1 \right),
\end{equation}
where instruction types are sorted in descending order of latency, i.e., 
$\ell_{(1)} \geq \ell_{(2)} \geq \cdots$, and 
$S_{(i)} = \sum_{k=1}^i n_{(k)}$ is the cumulative number of loads from the first $i$ operand types.
Intuitively, the formula assumes all loads are independent and can be reordered arbitrarily. 
The longest-latency loads are issued first so their delay is hidden as much as possible, while shorter-latency loads can be issued later and overlap with them. 
The final time is then determined by whichever group of loads (long or short) finishes last.
In Fig.~\ref{fig:accum_chain}, with 3 instruction types of latency $\ell=3$, counts $n_C=2$, $n_A=n_B=1$, and $U_{\text{LD}}=2$, we have $S_{(1)}=2$, $S_{(2)}=3$, $S_{(3)}=4$. 
Substituting gives $T_{\text{load}}=\max(3+\lceil 2/2\rceil-1,\;3+\lceil 3/2\rceil-1,\;3+\lceil 4/2\rceil-1)=\max(3,4,4)=4$ cycles.

\item \textbf{Steady State:} The region where \texttt{VMAC} instructions are interleaved with operand loads and scheduled periodically with an \emph{initiation interval} of $II$ cycles. 
For a single accumulation chain, $II$ is constrained by the RAW dependency distance on the accumulator operand and the load bandwidth, giving
\vspace{-3pt}
\begin{equation}
II_{\text{single}} = \max\!\left(P,\; \Big\lceil \tfrac{R_{\text{load}}}{U_{\text{LD}}} \Big\rceil \right),
\end{equation}
where $P$ is the \texttt{VMAC} pipeline depth, and $R_{\text{load}}$ is the number of operand loads per \texttt{VMAC}.
To hide the latency $P$ and improve \texttt{VMAC} throughput, a microkernel typically employs multiple parallel accumulation chains interleaved in a software-pipelined schedule. 
With $C$ parallel chains interleaved, the effective $II$ is: 
\vspace{-5pt}
\begin{equation}
II_{\text{parallel}} =\tfrac{\max \!\left(P + 1 - C,\; \Big\lceil \tfrac{R_{\text{load}}}{U_{\text{LD}}} \Big\rceil \right)}{C}.
\label{eq:II_parallel}
\end{equation}

The steady-state cost is then lower-bounded by:
\begin{equation}
\label{eq:T_steady}
T_{\text{steady}} \;\geq\; II_{\text{parallel}} \cdot  \big(N_{\text{accum}} - C),
\end{equation}
where $N_{\text{accum}}$ is the number of accumulator updates (i.e., the length of the accumulation chain).

\item \textbf{Epilog:} The region after the last operand load where the final \texttt{VMAC} instructions complete and the accumulator results are written back to memory. 
For $C$ accumulation chains, it consists of $C$ compute--store pairs:
\begin{equation}
T_{\text{epilog}} \;\geq\; \big(\ell_{\text{vmac}\rightarrow\text{vst}} \;+\; \ell_{\text{store}} \;+\; n_{\text{store}}-1\big)\;+\; (C-1),
\end{equation}
where $\ell_{\text{vmac}\rightarrow\text{vst}}$ is the latency from the last \texttt{VMAC} to when its store can be issued, 
$\ell_{\text{store}}$ is the store instruction latency, 
and $n_{\text{store}}$ is the number of store instructions required per accumulator.

\end{itemize}

\subsection{Modeling of Total Latency}

\paragraph{Double Buffering of Input Registers}
\begin{figure}[t]
  \centering
  \includegraphics[width=1\linewidth]{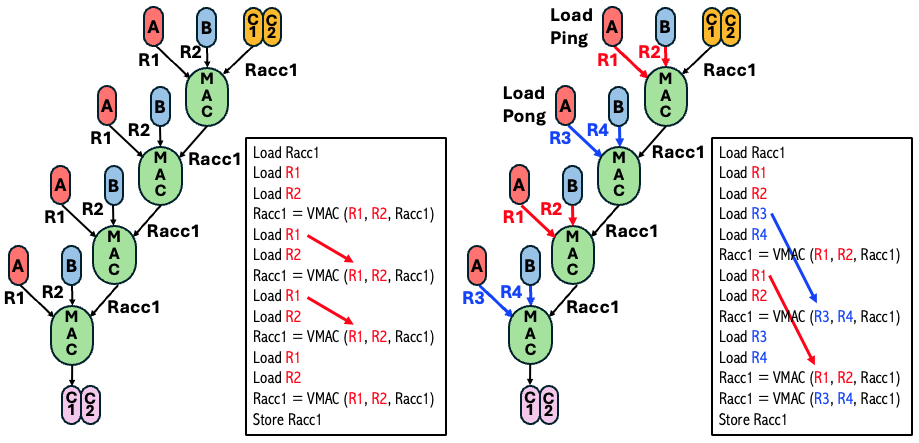}
  \vspace{-15pt}
  \caption{Comparison of single-buffered (left) vs. double-buffered (right) input register allocation.}
  \vspace{-15pt}
  \label{fig:double_buffering}
\end{figure}

In our steady-state modeling, we considered only the load bandwidth and assumed that sufficient registers are available to hold parallel loads, fully hiding the load latency. 
This assumption is reasonable in practice, as compilers often perform register renaming and allocate disjoint registers to achieve this effect automatically. 
When the compiler cannot apply such optimizations effectively, AIE programmers can manually pin variables to registers using the \texttt{chess\_storage(register)} attribute.
In this case, ping–pong buffering of input registers becomes an important technique. 
By alternating between two disjoint register sets (\emph{Ping}/\emph{Pong}), the microkernel can preload operands in one set while the other is consumed by \texttt{VMAC} instructions. 
Fig.~\ref{fig:double_buffering} illustrates this: the left schedule repeatedly reloads into the same registers, creating RAW dependencies between loads and compute, while the right employs double buffering so that loads overlap with compute.

\begin{figure}[t]
  \centering
  \includegraphics[width=\linewidth]{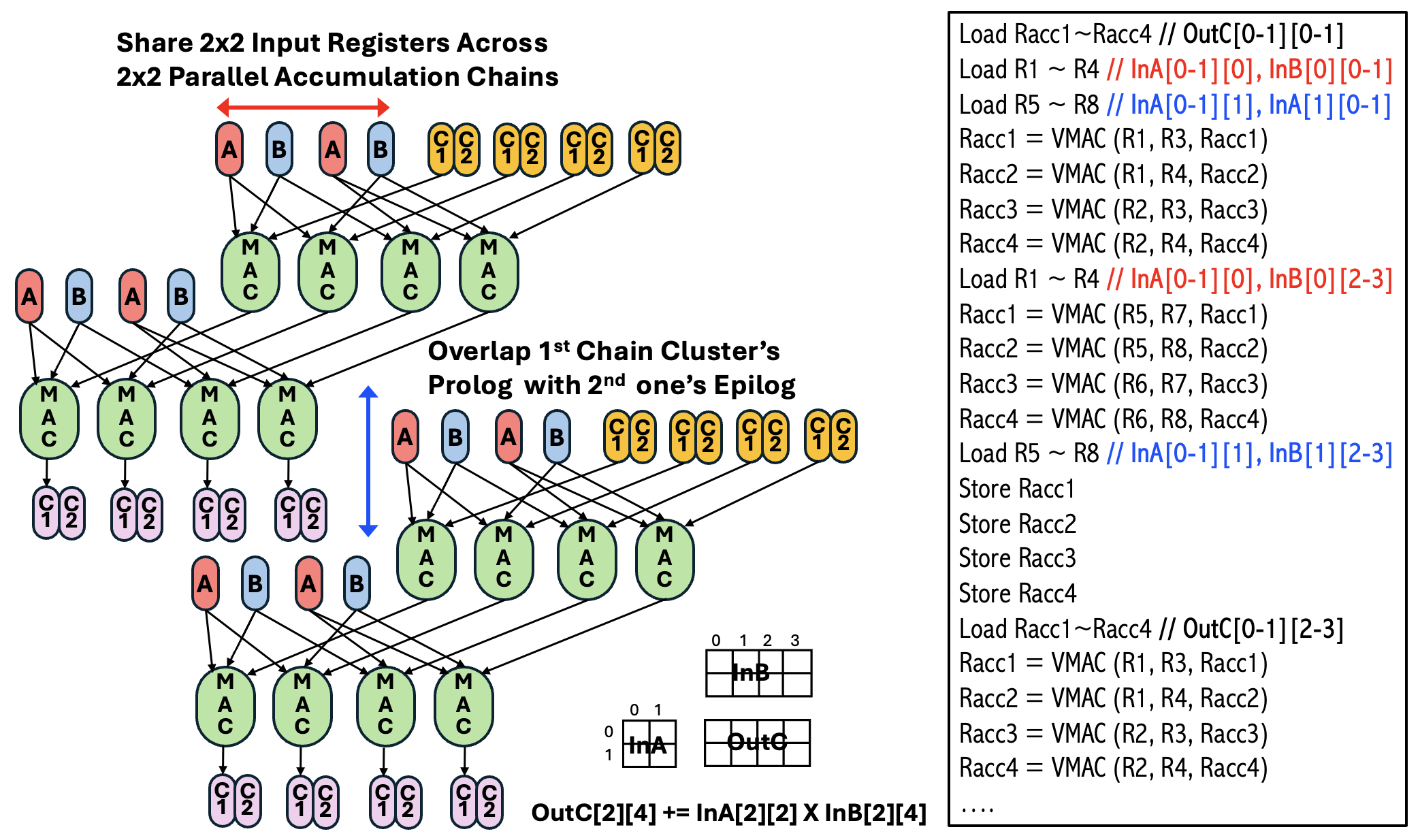}
\caption{GEMM microkernel optimizations: top-left, input sharing across a 2$\times$2 chain cluster; bottom-left, prolog–epilog overlap across clusters; right, corresponding code.}
~\label{fig:reuse_overlap}
    \vspace{-15pt}
\end{figure}

\paragraph{Sharing Input Across Chain Cluster}
To improve the microkernel performance, it is important to employ multiple parallel accumulation chains so that the effective $II$ in the steady state approaches~1, as modeled in Eq.~(\ref{eq:II_parallel}). 
In practice, these chains are not fully independent: they often share common input operands from matrix $A$ or $B$, which we refer to as a \emph{chain cluster}. 
When the microkernel shares input registers across a chain cluster, the number of loads per \texttt{VMAC} is reduced, thus alleviating pressure on the load bandwidth and lowering $R_{\text{load}}$ in Eq.~(\ref{eq:II_parallel}). 
This reuse also reduces input register usage, creating more opportunities for double buffering to hide load latency. 
The cluster size is bounded by the number of accumulator operands that can be held simultaneously in the accumulator register file. 
Fig.~\ref{fig:reuse_overlap} (top-left) illustrates this optimization, where shared inputs feed multiple chains simultaneously; for example, a 2$\times$2 chain cluster reduces the load count from 8 to 4.

\paragraph{Overlapping Successive Chain Clusters}
Even if the optimal $II=1$ is achieved in steady state using a chain cluster, the total latency still includes the prolog and epilog phases. 
If successive chain clusters run sequentially, with each starting only after the previous completes, the total latency is bounded by
\begin{equation}
\label{l_total_non_overlap}
L_{total} \;\geq\; (T_{\text{prolog}} + T_{\text{steady}} + T_{\text{epilog}}) \cdot N_{\text{clusters}}.
\end{equation}
However, the chain length is algorithmically constrained by the $K$ dimension of the tile in L1 memory. 
When $K$ is small, the prolog and epilog overheads are significant relative to $T_{\text{steady}}$.
To address this, successive clusters can overlap in a software-pipelined schedule: while one cluster is draining its epilog, the next cluster can begin its prolog. 
If the overlap ideally hides the boundary costs, 
the latency of $N_{\text{clusters}}$ chain clusters, each containing $C$ chains, is bounded by
\begin{equation}
\label{l_total_overlap}
L_{\text{total}} \;\geq\; T_{\text{prolog}} + \big(T_{\text{steady}} + II_{\text{parallel}} \cdot C\big) \cdot N_{\text{clusters}} + T_{\text{epilog}}.
\end{equation}
Fig.~\ref{fig:reuse_overlap} (bottom-left) illustrates this effect, where overlapping reduces the gaps between chain clusters for higher utilization.

\subsection{Towards BFP16 and Mixed Precision}

Since public documentation does not provide precise register behavior, we approximate parameters using microbenchmarks with small \texttt{VMAC} kernels and explicit register binding. For the \texttt{bfp16ebs8} \texttt{VMAC} instruction, we observe five accumulator registers, each holding $8 \times 8$ FP32 elements, with a pipeline distance of $P=3$ cycles. Operand loads have a latency of 8 cycles; because \texttt{bfp16ebs8} is unaligned, a \texttt{VLDA.POP} precedes the load whenever the access is to a non-consecutive address. The latency from \texttt{VMAC} to store availability is $\ell_{\text{vmac}\rightarrow\text{vst}}=6$ cycles, and writing back one accumulator register requires two store instructions with latency $\ell_{\text{store}}=2$ cycles.

Architectural constraints can reduce BFP16 kernel performance below the analytical bounds. The instruction pipeline favors sequential L1 access, because unaligned BFP16 addresses require extra instructions and may introduce bubbles. In addition, conversions between BFP16 and BF16 occupy accumulator registers, and the small accumulator file further limits concurrent load–store operations, which creates false dependencies between successive clusters.
Our model helps designers reason about whether performance loss stems from hardware limitations or compiler scheduling, and guides manual optimizations such as unrolling, register-level tiling, register binding, and double buffering.

\section{Tiling Tradeoffs Towards Final Performance}
~\label{sec:microkernel_model}
The final performance of the array is jointly bounded by
\emph{data movement} and \emph{compute capability}:
\begin{equation}
\label{eq:finalperf}
\text{Perf}_{\text{array}} \;=\;
\min\!\Big( \mathrm{AI}_{\text{array}} \cdot BW_{\text{offchip}}, \;\;
\text{Eff}_{\text{core}} \cdot 
\text{Perf}_{\text{core}}^{\text{peak}} \cdot 
N_{\text{core}} 
\Big),
\end{equation}
where $BW_{\text{offchip}}$ is the available off-chip bandwidth, which is around 65 GB/s for the Strix Point laptop.
$N_{\text{core}}=32$ is the number of compute cores,
$\text{Perf}_{\text{core}}^{\text{peak}}=1.84$  TFLOPS is the per-core peak performance, and
$\text{Eff}_{\text{core}}$ is the compute core efficiency.

\emph{Impact of $T^{L_1}_K$.}
A larger $T^{L_1}_K$ increases the accumulation chain length $N_\text{accum}$ in the microkernel,
allowing a longer steady state in Eq.~\ref{l_total_overlap} or Eq.~\ref{l_total_non_overlap}.
We use Eq.~\ref{l_total_non_overlap} for conservativeness.   
Combining it with Eq.~\ref{eq:T_steady}, 
we model the microkernel efficiency $\text{Eff}_{\text{micro}}$ as
\[
\label{eq:eff-microkernel}
\text{Eff}_{\text{micro}}
= \frac{N_{\text{clusters}} \cdot N_{\text{accum}}}
       {\bigl(T_{\text{prolog}} + II\cdot(N_{\text{accum}}-C) + T_{\text{epilog}}\bigr)\cdot N_{\text{clusters}}}.
\]
Since $N_{\text{accum}}$ scales with $T^{L_1}_K$, we can approximate
\begin{equation}
\text{Eff}_{\text{micro}}(T^{L_1}_K) 
\; =\; \tfrac{\eta\,T^{L_1}_K}{\epsilon + \eta\,II\,T^{L_1}_K}.
\end{equation}
This expression captures diminishing returns: $\text{Eff}_{\text{micro}}$ grows linearly 
for $T^{L_1}_K \!\ll\! \frac{\epsilon}{\eta\,II}$, but saturates to   $\frac{1}{II}$ as $T^{L_1}_K$ increases.

\emph{Impact of asymmetry.}
Between consecutive microkernels, we observe a fixed overhead 
for switching microkernels, measured as $\delta \!\approx\! 50$ cycles 
in tracing. 
For a core to finish an output tile $(T^{L_1}_{M_C}T^{L_1}_{N})$, in the symmetric case it requires $\tfrac{K}{T^{L_1}_K}$ switches, and $\rho \cdot \tfrac{K}{T^{L_1}_K}$ switches when using asymmetric tile buffering. Its latency is
\begin{equation}
\label{eq:latency-asym}
T_{\text{asym}}
= 
\tfrac{2 \, T^{L_1}_{M_C} \, K \, T^{L_1}_{N}}
     {\text{Perf}^{\text{peak}}_{\text{core}} \cdot \text{Eff}_{\text{micro}}}
\;+\;
\delta \cdot \rho \cdot \tfrac{K}{T^{L_1}_K}.
\end{equation}

\paragraph{Combined efficiency.}
The resulting effective core efficiency is
\begin{equation}
\label{eq:eff-core}
\text{Eff}_{\text{core}}
=
\frac{
2 \, T^{L_1}_{M_C} \, K \, T^{L_1}_{N}
}{
T_{\text{asym}}
}
\;=\;
\frac{
1
}{
\frac{1}{\text{Eff}_{\text{micro}}}
+
\frac{
\delta \, \rho \, \text{Perf}^{\text{peak}}_{\text{core}}
}{
2 \, T^{L_1}_{M_C} \, T^{L_1}_{N} \, T^{L_1}_K
}
}.
\end{equation}

The optimal strategy to maximize $\text{Eff}_{\text{core}}$ is to choose \textbf{$T_K$ as large as possible} and \textbf{$\rho$ as small as possible}, which is the opposite of the strategy that maximizes $\mathrm{AI}_\rho$. 
The final performance (Eq.~\ref{eq:finalperf}) therefore emerges from a \emph{tradeoff between maximizing arithmetic intensity and sustaining high compute efficiency}.

\section{Evaluation}


\begin{table}[t]
    \caption{Microkernel-level and core-level performance under different $T_K$ and $\rho$.}
    \label{tab:micro-core-eff}
    \centering
    \footnotesize
    \begin{tabular}{c c c c c c}
        \toprule
        $T_K$ & $\rho$ & Perf$_{\text{micro}}$ (Tflops) & Eff$_{\text{micro}}$ & Perf$_{\text{core}}$ (Tflops)& Eff$_{\text{core}}$ \\
        \midrule
         8  & 1 & 0.36       &  0.2      & 0.290 & 0.156 \\
         8  & 2 & 0.36       &  0.2     & 0.277 & 0.149 \\
         8  & 4 & 0.36       &  0.2      & 0.250 & 0.134 \\
         8  & 8 & 0.36       &  0.2      & 0.216 & 0.116  \\ \midrule
        16  & 1 & 0.66       &  0.36      & 0.529 & 0.284 \\ 
        16  & 2 & 0.66       &   0.36     &  0.501     &  0.272     \\
        16  & 4 & 0.66  &   0.36     &    0.474   &   0.257    \\
        16  & 8 & 0.66       &   0.36     &   0.359    &   0.195   \\ \midrule
        32  & 2 & 0.76       &   0.41     & 0.615 & 0.331 \\
        32  & 4 & 0.75  &   0.41     &   0.574    &   0.312    \\
        32  & 8 & 0.77       &    0.42    &  0.562     &  0.306      \\ \midrule
        64  & 4 & 1.16       &   0.63     & 0.920 & 0.511 \\
        64  & 8 & 1.16     &  0.63      & 0.867 & 0.482 \\
        \bottomrule
    \end{tabular}
\end{table}

\begin{table*}[t]
\renewcommand{\arraystretch}{0.7}
\caption{Whole-array BF16/BFP16 GEMM speedup across configurations.}
\footnotesize
\label{tab:gemm-speedup-final}
\centering
\begin{tabular}{llcccccccccc}
\toprule
\makecell{Precision} & \makecell{Source}
  & \makecell{Problem size\\(M$\times$K$\times$N)}
  & \makecell{L1 tile size\\($T^{L_1}_{M_C}$$\times$$T^{L_1}_K$$\times$$T^{L_1}_N$)}
  & \makecell{$\rho$}
  & \makecell{Used Buffer \\ (KB)}
  & \makecell{Buffer if \\ $\rho=1$}  
  & \makecell{$\text{Perf}_{\text{core}}$\\ (Tflops)}
  & \makecell{Compute-bound \\ (Tflops)}
  & \makecell{$\mathrm{AI}_{\text{array}}$ \\ (op/B)}
  & \makecell{Memory-bound \\ (Tflops)}
  & \makecell{ $\text{Perf}_{\text{array}}$\\(Tflops)} 
\\
\midrule
\makecell[l]{BF16-INT4}
  & Ryzen AI & 2048$\times$4096$\times$4096 & Unknown & 1 & Unk & Unk & Unk & Unk & Unk & Unk & 10.4 \\
\midrule
\multirow{5}{*}{\makecell[l]{Config 1}}
  & MLIR-AIE & 8192$\times$4224$\times$4096 & 64$\times$88$\times$64 & 1 & 50.4 & 50.4 & 0.32 & \textbf{10.2} & 216 & 14.1 & 4.8 (1.00$\times$) \\
  \cmidrule(lr){2-12}
  & \multirow{3}{*}{Ours} & 2048$\times$4096$\times$2048 & 64$\times$64$\times$128   & 1 &54.5&  54.5 & 0.95 & 30.5 & 273 & \textbf{17.8} & 17.3 (3.61$\times$) \\
  &                       & 2048$\times$4480$\times$2048 & 64$\times$224$\times$64   & 4 &57.4& 111 & 0.96 & 30.7 & 217 & \textbf{14.1} & 12.3 (2.56$\times$) \\
  &                       & 4096$\times$4096$\times$2048 & 128$\times$64$\times$128   & 4 &60.3& 91 & 0.95 & 30.4 & 410 & \textbf{26.6} & 24.3 (4.54$\times$) \\
\midrule
\multirow{4}{*}{\makecell[l]{Config 2}}
  & MLIR-AIE & 7680$\times$4096$\times$8192 & 96$\times$128$\times$64  & 1 & 58.5 & 58.5 & 0.38 & \textbf{12.2} & 333 & 21.7 & 10.2 (1.00$\times$) \\
  \cmidrule(lr){2-12}
  & \multirow{2}{*}{Ours} & 4096$\times$4096$\times$2048 & 128$\times$64$\times$128  & 1 & 54 & 54 & 0.84 & \textbf{26.9} & 504 & 32.8 & 25.9 (2.54$\times$) \\ 
  &                       & 4096$\times$4096$\times$2048 & 256$\times$64$\times$128 & 8 & 58.5 & 90 & 0.73 & \textbf{23.3} & 728 & 47.3 & 22.3 (2.19$\times$)  \\
  &                      & 3072$\times$4096$\times$1536 & 192$\times$128$\times$96 & 6 & 56.3 & 101 & 1.06 & \textbf{33.9} & 562 & 36.5  & 31.3 (3.07$\times$) \\
\midrule
\makecell[l]{Config 3}
  &       Ours           & 3072$\times$4096$\times$2048              & 96$\times$64$\times$128  & 1 & 57 &  57 & 1.00 & 32 & 418 & \textbf{27.2} & 27.0 (2.65$\times$) \\
  &                & 4096$\times$4096$\times$2048              & 128$\times$64$\times$128  & 4 & 56.5 &  70 & 0.90 & \textbf{28.8} & 504 & 32.8 & 28.5 (2.79$\times$) \\
\bottomrule
\end{tabular}
\end{table*}

\subsection{Experimental Setup}
Our experiments use AMD Ryzen\texttrademark{} AI 9 HX 370 (Strix Point NPU). 
We program the NPU using the IRON API and the MLIR-AIE toolchain~\cite{fccmIRON} to measure performance under three precision configurations:
\textbf{Config 1:} BF16--BFP16 mixed-precision. Input $A$ and output $C$ are in BF16, weights $B$ are in BFP16, and accumulation is performed in BF16.
\textbf{Config 2:} $A$, $B$, and $C$ are all stored in BFP16, with accumulation also performed in BFP16.
\textbf{Config 3:} $A$, $B$, and $C$ are all stored in BFP16, with accumulation performed in BF16.



\subsection{Kernel Performance}

Table~\ref{tab:micro-core-eff} summarizes both microkernel-level performance 
(Perf$_{\text{micro}}$, Eff$_{\text{micro}}$) and full core-level performance 
(Perf$_{\text{core}}$, Eff$_{\text{core}}$) across different combinations of 
$T_K$ and $\rho$, evaluated under precision Config~1 with L1 tile sizes 
$T_{M_A} = 128$ and $T_N = 128$.
For $T_K = 64$, the configurations 
$\rho = 1$ and $\rho = 2$ exceed the available 63 KB L1 memory capacity, and for 
$T_K = 32$, $\rho = 1$ also exceeds it.

Overall, the measured results closely follow the analytical trends predicted in 
Section~\ref{sec:microkernel_model}. As expected, microkernel efficiency Eff$_{\text{micro}}$ depends only on $T_K$ and is almost invariant across $\rho$. For a fixed $T_K$, increasing $\rho$ 
consistently reduces core efficiency Eff$_{\text{core}}$ because larger asymmetry increases the 
number of microkernel switches and amplifies launch overheads. 
For example, at $T_K=8$, Eff$_{\text{core}}$ decreases by 25.6\% (from $0.156$ to $0.116$) as $\rho$ increases from $1$ to $8$. At $T_K=64$, the reduction is much smaller, only 5.8\% (from $0.511$ to $0.482$), because a larger $T_K$ yields longer kernel execution and therefore better amortization of the switching cost.
In contrast, for a fixed $\rho$, a larger $T_K$ improves both the efficiency of the microkernel and the core by extending the steady state and reducing the relative impact of the prolog/epilog cycles. 
At $\rho=4$, Eff$_{\text{core}}$ increases 
from $0.134$ at $T_K=8$ to $0.511$ at $T_K=64$, a $3.8\times$ improvement.
These results match the theoretical prediction that core efficiency improves with larger $T_K$ while increasing $\rho$ reduces efficiency.

Additionally, MLIR-AIE \cite{mlir-aie-matmul} achieves only 0.32 TFLOPS on a single core, while our best design reaches 0.92 TFLOPS—a 2.88× improvement. Their approach relies primarily on compiler-driven optimization, which is ineffective for BFP16, while our kernel design applies manual scheduling techniques such as unrolling, register-level tiling, explicit register binding, and double buffering.

\begin{table}[htpb]
\renewcommand{\arraystretch}{1} 
\caption{Performance of whole-array GEMM under varying matrix problem sizes.}
\label{tab:gemm-scaling}

\centering
\footnotesize
\begin{tabular}{cccc}

\toprule
M & K & N & \makecell{Performance (Tflops)} \\
\midrule
1024 & 2048 & 2048 & 19.2 \\
1024 & 4096 & 2048 & 21.8 \\
1024 & 8192 & 2048 & 23.3 \\
2048 & 2048 & 2048 & 21.8 \\
2048 & 4096 & 2048 & 23.3 \\
2048 & 8192 & 2048 & 24.0 \\
4096 & 2048 & 2048 & 23.2 \\
4096 & 4096 & 2048 & 24.3 \\
4096 & 8192 & 2048 & 24.6 \\
\bottomrule
\end{tabular}
\end{table}


\subsection{Whole Array Performance}
Table~\ref{tab:gemm-speedup-final} reports full-array GEMM throughput across
three configurations, comparing our kernels against both the AMD MLIR-AIE \cite{mlir-aie-matmul} and the BF16--INT4 backend used in AMD Ryzen~AI~\cite{Ryzen-AI}
(DynamicDispatch~\cite{DynamicDispatch}). For each problem size, the table
lists the measured single-core throughput, the compute-bound limit
(Perf$_{\text{core}} \times 32$ cores), the arithmetic intensity, and
the corresponding memory-bound limit ( $\mathrm{AI}_{\text{array}}$~$\times$~off-chip bandwidth), as
defined in Eq.~\ref{eq:finalperf}.

Across all configurations, our kernels substantially outperform the
state-of-the-art MLIR-AIE and Ryzen~AI baselines. In Config~1, our best design reaches
24.3~TFLOPS, a 4.54$\times$ speedup over MLIR-AIE’s 4.8~TFLOPS. Config~2
shows similar gains, with throughput up to 31.3~TFLOPS (3.07$\times$ over
the 10.2~TFLOPS baseline). Config~3 achieves 28.5~TFLOPS, exceeding the
MLIR-AIE Config~2 by 2.79$\times$ while simultaneously offering higher accumulation precision.
Compared to the Ryzen~AI BF16–INT4 GEMM (10.4~TFLOPS), our Config~1
BF16–BFP16 GEMM delivers 24.3~TFLOPS, providing a 2.4$\times$ improvement
while using higher precision.

We can observe the impact of ATB in 
Table~\ref{tab:gemm-speedup-final}. In Config~1, all three configurations are memory-bound. 
ATB enables the $128\!\times\!64\!\times\!128$ tile, which would require 
91~KB and be infeasible under symmetric buffering. With ATB, this configuration 
becomes valid and improves throughput from 17.3 to 24.3~TFLOPS—a 40\% gain. 
Fixing $\rho=4$ also highlights the effect of $T_K$: the 
$64\!\times\!224\!\times\!64$ kernel achieves only 56\% of the throughput 
of the $128\!\times\!64\!\times\!128$ tile, since larger $T_K$ forces 
smaller $(T_{M_C},T_N)$ and reduces AI. These observations match the 
theoretical prediction that memory-bound kernels benefit from smaller $T_K$ 
and larger $\rho$, which enable larger $(T_{M_C},T_N)$.

Config~2 is mostly compute-bound and shows the trade-off between AI and core 
efficiency. Although $256\!\times\!64\!\times\!128$ has higher AI, its 
lower core efficiency limits performance; the 
$192\!\times\!128\!\times\!96$ tile, with larger $T_K$, reaches 
31.3~TFLOPS. ATB is again essential: without it, this configuration would 
require 101~KB and would not fit in L1, and ATB provides a 21\% throughput 
gain over the symmetric case.
Config~3 further shows that ATB can move GEMM from memory-bound to 
compute-bound. The final performance does not increase noticeably because the kernel is already compute-bound.
In summary, when memory-bound, ATB helps increase arithmetic intensity through larger $(T_{M_C},T_N)$; when compute-bound, ATB helps improve core efficiency through larger $T_K$.

Finally, Table~\ref{tab:gemm-scaling} reports the performance of our Config~1 ATB kernel ($\rho=4$, $128\times64\times128$ L1 tile) across different problem dimensions. Performance scales smoothly with matrix size, achieving up to 24.6~TFLOPS on large GEMMs and remaining above 19.2~TFLOPS for smaller problems.

\section{Conclusion}
In this work, we introduced asymmetric tile buffering (ATB), a general tiling strategy that reduces buffer pressure and improves arithmetic intensity beyond conventional symmetric buffering on NPUs. On AMD XDNA2™ AIE, ATB improves GEMM throughput by up to 40\% even over highly optimized symmetric kernels. We also present kernel design strategies that deliver a 4.54× speedup over the state-of-the-art GEMM implementation.
In the future, we plan to explore automated search and scheduling to make ATB-based kernel generation fully automatic.

\bibliographystyle{ACM-Reference-Format}
\bibliography{references}

@inproceedings{rsn,
author = {Wang, Chengyue and Zhang, Xiaofan and Cong, Jason and Hoe, James C.},
title = {Reconfigurable Stream Network Architecture},
year = {2025},
isbn = {9798400712616},
publisher = {Association for Computing Machinery},
address = {New York, NY, USA},
url = {https://doi.org/10.1145/3695053.3731088},
doi = {10.1145/3695053.3731088},
booktitle = {Proceedings of the 52nd Annual International Symposium on Computer Architecture},
pages = {1848–1866},
numpages = {19},
location = {
},
series = {ISCA '25}
}

@misc{mhatre2025gamahighperformancegemmacceleration,
      title={GAMA: High-Performance GEMM Acceleration on AMD Versal ML-Optimized AI Engines}, 
      author={Kaustubh Mhatre and Endri Taka and Aman Arora},
      year={2025},
      eprint={2504.09688},
      archivePrefix={arXiv},
      primaryClass={cs.AR},
      url={https://arxiv.org/abs/2504.09688}, 
}

@misc{taka2023maxevamaximizingefficiencymatrix,
      title={MaxEVA: Maximizing the Efficiency of Matrix Multiplication on Versal AI Engine}, 
      author={Endri Taka and Aman Arora and Kai-Chiang Wu and Diana Marculescu},
      year={2023},
      eprint={2311.04980},
      archivePrefix={arXiv},
      primaryClass={cs.AR},
      url={https://arxiv.org/abs/2311.04980}, 
}

@INPROCEEDINGS{fccmIRON,
  author={Hunhoff, Erika and Melber, Joseph and Denolf, Kristof and Bisca, Andra and Bayliss, Samuel and Neuendorffer, Stephen and Fifield, Jeff and Lo, Jack and Vasireddy, Pranathi and James-Roxby, Phil and Keller, Eric},
  booktitle={2025 IEEE 33rd Annual International Symposium on Field-Programmable Custom Computing Machines (FCCM)}, 
  title={Efficiency, Expressivity, and Extensibility in a Close-to-Metal NPU Programming Interface}, 
  year={2025},
  volume={},
  number={},
  pages={85-94},
  doi={10.1109/FCCM62733.2025.00043}}

@INPROCEEDINGS{AMA_FPL_2024,
  author={Deng, Xiaodong and Wang, Shijie and Gao, Tianyi and Liu, Jing and Liu, Longjun and Zheng, Nanning},
  booktitle={2024 34th International Conference on Field-Programmable Logic and Applications (FPL)}, 
  title={{AMA: An Analytical Approach to Maximizing the Efficiency of Deep Learning on Versal AI Engine}}, 
  year={2024},
  volume={},
  number={},
  pages={227-235},
  doi={10.1109/FPL64840.2024.00039}}

@INPROCEEDINGS{Versal_vs_Stratix_FCCM_2024,
  author={Taka, Endri and Gourounas, Dimitrios and Gerstlauer, Andreas and Marculescu, Diana and Arora, Aman},
  booktitle={2024 IEEE 32nd Annual International Symposium on Field-Programmable Custom Computing Machines (FCCM)}, 
  title={{Efficient Approaches for GEMM Acceleration on Leading AI-Optimized FPGAs}}, 
  year={2024},
  volume={},
  number={},
  pages={54-65},
  doi={10.1109/FCCM60383.2024.00015}}

@inproceedings{Gillies-Liu-ANDOR-1990,
  author    = {D. W. Gillies and J. W.-S. Liu},
  title     = {Scheduling Tasks with AND/OR Precedence Constraints},
  booktitle = {Proceedings of the Second IEEE Symposium on Parallel and Distributed Processing},
  year      = {1990},
  pages     = {394--401},
  doi       = {10.1109/SPDP.1990.143572}
}

@misc{zhuang2023charmcomposingheterogeneousaccelerators,
      title={CHARM: Composing Heterogeneous Accelerators for Matrix Multiply on Versal ACAP Architecture}, 
      author={Jinming Zhuang and Jason Lau and Hanchen Ye and Zhuoping Yang and Yubo Du and Jack Lo and Kristof Denolf and Stephen Neuendorffer and Alex Jones and Jingtong Hu and Deming Chen and Jason Cong and Peipei Zhou},
      year={2023},
      eprint={2301.02359},
      archivePrefix={arXiv},
      primaryClass={cs.AR},
      url={https://arxiv.org/abs/2301.02359}, 
}

@misc{zhuang2023autommenergyefficientmultidatatypematrix,
      title={AutoMM: Energy-Efficient Multi-Data-Type Matrix Multiply Design on Heterogeneous Programmable System-on-Chip}, 
      author={Jinming Zhuang and Zhuoping Yang and Peipei Zhou},
      year={2023},
      eprint={2305.18698},
      archivePrefix={arXiv},
      primaryClass={cs.AR},
      url={https://arxiv.org/abs/2305.18698}, 
}

@misc{mei2020zigzagmemorycentricrapiddnn,
      title={ZigZag: A Memory-Centric Rapid DNN Accelerator Design Space Exploration Framework}, 
      author={Linyan Mei and Pouya Houshmand and Vikram Jain and Sebastian Giraldo and Marian Verhelst},
      year={2020},
      eprint={2007.11360},
      archivePrefix={arXiv},
      primaryClass={cs.DC},
      url={https://arxiv.org/abs/2007.11360}, 
}

@article{gilbert2024looptree,
    author = {Gilbert, Michael and Wu, Yannan and Emer, Joel and Sze, Vivienne},
    year = {2024},
    month = {09},
    pages = {1-15},
    title = {LoopTree: Exploring the Fused-Layer Dataflow Accelerator Design Space},
    volume = {PP},
    journal = {IEEE Transactions on Circuits and Systems for Artificial Intelligence},
    doi = {10.1109/TCASAI.2024.3461716}
}

@software{Thakkar_CUTLASS_2023,
    author = {Thakkar, Vijay and Ramani, Pradeep and Cecka, Cris and Shivam, Aniket and Lu, Honghao and Yan, Ethan and Kosaian, Jack and Hoemmen, Mark and Wu, Haicheng and Kerr, Andrew and Nicely, Matt and Merrill, Duane and Blasig, Dustyn and Qiao, Fengqi and Majcher, Piotr and Springer, Paul and Hohnerbach, Markus and Wang, Jin and Gupta, Manish},
    license = {BSD-3-Clause},
    month = jan,
    title = {{CUTLASS}},
    url = {https://github.com/NVIDIA/cutlass},
    version = {3.0.0},
    year = {2023}
}

@article{BLIS1,
   author      = {Field G. {V}an~{Z}ee and Robert A. {v}an~{d}e~{G}eijn},
   title       = {{BLIS}: A Framework for Rapidly Instantiating {BLAS} Functionality},
   journal     = {ACM Transactions on Mathematical Software},
   volume      = {41},
   number      = {3},
   pages       = {14:1--14:33},
   month       = {June},
   year        = {2015},
   issue_date  = {June 2015},
   url         = {https://doi.acm.org/10.1145/2764454},
}

@misc{mlir-aie-matmul,
  author = {AMD},
  year = {2025},
  title        = {Xilinx MLIR-AIE BFP16 matrix multiplication implementation},
  howpublished = {\url{https://github.com/Xilinx/mlir-aie/tree/main/programming_examples/ml/block_datatypes/matrix_multiplication}},
  note         = {Accessed: 2025-10-01}
}

@misc{amd-bfp16-quark,
  author       = {AMD},
  year         = {2025},
  title        = {BFP16 (Block Floating Point) Quantization Tutorial},
  howpublished = {\url{https://quark.docs.amd.com/latest/pytorch/tutorial_bfp16.html}},
  note         = {Accessed: 2025-10-01}
}

@misc{amd-aie-api-mmul,
  author       = {AMD},
  year         = {2025},
  title        = {AI Engine API – Matrix Multiplication (MMUL) Reference},
  howpublished = {\url{https://download.amd.com/docnav/aiengine/xilinx2025_1/aiengine_api/aie_api/doc/group__group__mmul.html}},
  note         = {Accessed: 2025-10-01}
}

@misc{Ryzen-AI,
  author = {AMD},
  year = {2025},
  title        = {Ryzen AI Software},
  howpublished = {\url{https://github.com/amd/RyzenAI-SW?tab=readme-ov-file}},
  note         = {Accessed: 2025-10-01}
}

@misc{DynamicDispatch,
  author = {AMD},
  year = {2025},
  title        = {DynamicDispatch},
  howpublished ={\url{https://github.com/amd/DynamicDispatch}},
  note         = {Accessed: 2025-10-01}
}

@misc{AMD-xdna-driver,
  author       = {AMD},
  year         = {2025},
  title        = {AMD XDNA Driver for Linux},
  howpublished = {\url{https://github.com/amd/xdna-driver}},
  note         = {Accessed: 2025-11-16}
}

@misc{amd_adaptive_overview,
  author       = {{AMD}},
    year         = {2023},
   title        = {Adaptive Computing Technology Overview},
  howpublished = {\url{https://www.amd.com/content/dam/amd/en/documents/solutions/adaptive-computing-technology-overview.pdf}},
  note         = {Accessed: 2025-11-19}
}

@INPROCEEDINGS{mindgap,
  author={Huang, Qijing and Tsai, Po-An and Emer, Joel S. and Parashar, Angshuman},
  booktitle={2024 ACM/IEEE 51st Annual International Symposium on Computer Architecture (ISCA)}, 
  title={Mind the Gap: Attainable Data Movement and Operational Intensity Bounds for Tensor Algorithms}, 
  year={2024},
  volume={},
  number={},
  pages={150-166},
  doi={10.1109/ISCA59077.2024.00021}}

@article{roofline,
author = {Williams, Samuel and Waterman, Andrew and Patterson, David},
title = {Roofline: an insightful visual performance model for multicore architectures},
year = {2009},
issue_date = {April 2009},
publisher = {Association for Computing Machinery},
address = {New York, NY, USA},
volume = {52},
number = {4},
issn = {0001-0782},
url = {https://doi.org/10.1145/1498765.1498785},
doi = {10.1145/1498765.1498785},
journal = {Commun. ACM},
month = apr,
pages = {65–76},
numpages = {12}
}

@misc{amd-ryzen-ai-9-hx-370,
  author       = {{AMD}},
  title        = {{AMD Ryzen AI 9 HX 370} Processor},
  year         = {2024},
  url          = {https://www.amd.com/en/products/processors/laptop/ryzen/ai-300-series/amd-ryzen-ai-9-hx-370.html},
  note         = {Accessed: 2025-11-18}
}

@misc{amd-ryzen-ai-5-340,
  author       = {{AMD}},
  title        = {{AMD Ryzen AI 5 340} Processor},
  year         = {2024},
  url          = {https://www.amd.com/en/products/processors/laptop/ryzen/ai-300-series/amd-ryzen-ai-5-340.html},
  note         = {Accessed: 2025-11-18}
}










\end{document}